\documentclass[amssymb,preprint,preprintnumbers,nofootinbib,superscriptaddress]{revtex4}
\usepackage{amsmath}
\usepackage{graphicx}
\graphicspath{{Figures/}}
\usepackage{latexsym}
\usepackage{amsfonts}
\usepackage{url,hyperref}
\usepackage{bm}
\usepackage{textcomp}
\usepackage{color}
\usepackage{bbm}
\usepackage{slashed}
\usepackage{caption}
\usepackage{epstopdf}
\usepackage{subcaption}
\usepackage{placeins}
\usepackage{color}
\usepackage{cancel} 
\usepackage[normalem]{ulem}
\usepackage{tcolorbox}
\usepackage{soul}

\begin{document}

\title{Gravitational lensing by traversable wormholes supported by three-form fields}

\author{Daris Samart}
\email{darisa@kku.ac.th}
\affiliation{Khon Kaen Particle Physics and Cosmology Theory Group (KKPaCT), Department of Physics, Faculty of Science, Khon Kaen University, Khon Kaen, 40002, Thailand}
\author{Natthason Autthisin}%
\email{natthasorn\_ut@kkumail.com}
\affiliation{Khon Kaen Particle Physics and Cosmology Theory Group (KKPaCT), Department of Physics, Faculty of Science, Khon Kaen University, Khon Kaen, 40002, Thailand}

\author{Phongpichit Channuie}
\email{channuie@gmail.com}
\affiliation{College of Graduate Studies, Walailak University, Thasala, Nakhon Si Thammarat, 80160, Thailand}
\affiliation{School of Science, Walailak University, Thasala, Nakhon Si Thammarat, 80160, Thailand}

\begin{abstract}

In this paper, we study the deflection angle of light by traversable wormholes, which are supported by the three-form fields. The specific forms of the redshift and shape functions that produce results compatible with the energy conditions at throat of the wormholes are used. Having used the well-defined parameter sets of the three-form wormholes, we investigate the photon geodesic motion under the effective potential of the wormhole background. As a result, we discover that the radius of the photon sphere is a very useful observable that is used to analyze the geometrical structures of a physical wormhole. 

\end{abstract}
\maketitle{}


\section{Introduction}\label{sec: intro}

Over the past few decades, wormholes have become
one of the most intensively studied topics in the literature. The wormhole solutions represent a conduit between the points of two parallel universes or even two different points of the same universe. The existence of traversable wormholes is known as Einstein-Rosen bridges proposed by Einstein and Rosen in 1935 \cite{Einstein:1935tc}. However, Wheeler and his colleague later showed that wormholes would not be stable and can not be traversable \cite{Wheeler:1955zz,Fuller:1962zza}. In 1988, Morris, Thorne, and Yurtsever have demonstrated that the traversable wormholes were explicitly constructed \cite{Morris:1988cz}. Later, other types of traversable wormholes were studied as plausible solutions to the equations of general relativity, including analyses initiated by Matt Visser \cite{Visser1995}

Searching for a proper form of the exotic matter inside the wormholes is a central topic in the wormhole research. In this work, we will use the the three-form fields as the exotic matter that can support the throat of the traversable wormholes. The three-form fields naturally exist in the string theory \cite{Ovrut:1997ur,Klebanov:2000hb,Bousso:2000xa,Frey:2002qc,Groh:2012tf,Bielleman:2015ina,Bandos:2012gz,Farakos:2017jme,Bandos:2018gjp,Farakos:2017ocw}. Moreover, it has been used to explain the several problems in cosmology such as inflation in very early universe and structure formation \cite{Koivisto:2009fb,Germani:2009gg,DeFelice:2012jt,DeFelice:2012wy,Kumar:2014oka}, and even the dark energy problem \cite{Koivisto:2009fb,Koivisto:2009ew,Ngampitipan:2011se,Koivisto:2012xm}. The traversable wormholes with the three-form fields were constructed by authors of Ref.\cite{Barros:2018lca}. It was shown that the three-form fields are an interesting exotic matter possibly used to support the wormholes throat. Consequently, the three-form fields are also used to study relativistic stars and black holes \cite{Barros:2020ghz,Bouhmadi-Lopez:2020wve,Barros:2021jbt}. Recently, the three-form wormholes are studied in detail by using the Higgs type three-form potential \cite{Bouhmadi-Lopez:2021zwt}.

To prove the existence of the wormholes, there are several methods that can be used to figure the wormholes. Gravitational lensing is one of the interesting phenomena which can be used to probe the astrophysical objects such as black holes. Gravitational lensing in wormholes has been explored in
many aspects and compared with the study of black holes \cite{Nandi:2006ds,Rahaman:2007am,Dey:2008kn,Bhattacharya:2010zzb,Abe:2010ap,Nakajima:2012pu,Sharif:2015qfa,Tsukamoto:2016qro,Tsukamoto:2016zdu,Nandi:2016uzg,Nandi:2018mzm}. Using the Gauss-Bonnet theorem, gravitational lensing by many types of black holes and wormholes have also been extensively studied, e.g., \cite{Jusufi:2017mav,Jusufi:2018kmk,Ovgun:2018prw}. The gravitational lensing tells us about the geometry of space-time around massive object and wormhole is one of them. Therefore, this is a promising idea to search for the existence of the wormholes. In addition, gravitational lensing occurs when the light travel passing through the space-time around the massive objects \cite{congdon2018book}. Consequently, this leads to the bending of light between source and observers according to the curve space-time around the massive object. One of the striking predictions of the gravitational lensing is the Einstein’s rings \cite{Shaikh:2019jfr}, commonly observed phenomena of the light bending in astrophysics.

In this work, we use the gravitational lensing as the main tool to search for the existence of the wormholes. Gravitational lensing is one of the classical predictions by GR. This phenomenon tells us about distribution of matter and manifests the curved space-time. This makes the curve geodesic line and leads to the bending of light between the light source and observers due to the massive object in the space-time \cite{Carroll:2004st,Hartle:2003yu}. The trajectory of the light passing through the wormholes will be curved by the space-time geometry of the wormholes. We will calculate and discuss a strong gravitational lensing and an Einstein’s rings which are very useful in the search of the traversable wormholes. 

The present work is organized as follows: In Sec.\ref{formalism}, all relevant equations of the three-form wormholes are briefly constructed and some implications are revisited. Next, in Sec.\ref{lens}, we set up the analytical expression of thee effective potential and deflection angle of the three-form wormholes. The numerical results are presented and its physical interpretations are given in section \ref{numer-results}. Finally, we conclude our findings in the last section \ref{conclusion}.

\section{Background equations with three-form fields}\label{formalism}

\subsection{Gravitational three-form fields action and spherical symmetric solution ansatz}
We start with the gravitational action of the Einstein gravity including the three-form fields. In this case, the action is of the form:
\begin{eqnarray}
S = \int \,d^4x\left[ \frac{1}{2\,\kappa^2}\,R -\frac{1}{48}\,F_{\mu\nu\rho\sigma}\,F^{\mu\nu\rho\sigma} - V(A^2) \right],
\label{action}
\end{eqnarray}
where $\kappa^2 \equiv 8\pi\,G$,\, $A^2 = A_{\mu\nu\rho}\,A^{\mu\nu\rho}$ and $R$ is a Ricci scalar. Additionally, $F_{\mu\nu\rho\sigma}$ is a field strength tensor of the three-form fields, $A_{\mu\nu\rho}$. It is defined as
\begin{eqnarray}
F_{\mu\nu\rho\sigma} = \nabla_\mu\,A_{\nu\rho\sigma} - \nabla_\sigma\,A_{\mu\nu\rho} + \nabla_\rho\,A_{\sigma\mu\nu\rho} - \nabla_\nu\,A_{\rho\sigma\mu}\,.
\label{3-form-stength}
\end{eqnarray}
Varying the action in (\ref{action}), the Einstein field equation is given by
\begin{eqnarray}
G_{\mu\nu} = \kappa^2\,T_{\mu\nu}\,,
\label{efe}
\end{eqnarray}
where $G_{\mu\nu} \equiv R_{\mu\nu} -\frac12\,g_{\mu\nu}\,R$\,. The energy-momentum tensor of the three-form fields reads
\begin{eqnarray}
T_{\mu\nu} \equiv \frac{1}{6}\,F_{\mu}^{~\,\rho\sigma\tau}\,F_{\nu\rho\sigma\tau} + 6\frac{\partial\,V(A^2)}{\partial (A^2)}\,A_{\mu}^{~\,\rho\sigma}\,A_{\nu\rho\sigma} - \left[\frac{1}{48}\,F_{\mu\nu\rho\sigma}\,F^{\mu\nu\rho\sigma} - V(A^2) \right]\,g_{\mu\nu}\,.
\label{EMT-3-form}
\end{eqnarray}
The equation of motion of the three-form is obtained by varying with respect to the $A_{\mu\nu\rho}$ and we find
\begin{eqnarray}
\nabla_\sigma F^{\sigma\mu\nu\rho} - 12\frac{\partial\,V(A^2)}{\partial (A^2)}\,A^{\mu\nu\rho} = 0\,.
\end{eqnarray}
The three-form gauge field, $A_{\mu\nu\rho}$ is represented by the following relation 
\begin{equation}
A_{\mu\nu\rho}=\sqrt{-g}\epsilon_{\mu\nu\rho\sigma}B^{\sigma}, \label{3-form-dual}   
\end{equation}
where the $B^\sigma$ is the one-form dual vector of the three-form fields, $A_{\mu\nu\rho}$.
The solution of the $B^\sigma$ can be expressed by using ansatz  as a function of $\zeta(r)$ via  \cite{Barros:2018lca}
\begin{equation}
B^{\sigma}=\left(0,\sqrt{1-\frac{b(r)}{r}}\zeta(r),0,0\right)^{\intercal}. \label{3-form-dual-comp}  
\end{equation}
where $\zeta(r)$ is a generic parametrization of the three-form fields and we will solve the EFE in Eq.(\ref{efe}) in order to obtain the solution of $\zeta(r)$.

Before imposing the ansatz of the three-from field solution, it is worth introducing the spherical symmetry line element of the traversable wormholes used in the present work. We consider a general spherically symmetric space-time geometry and static traversable wormholes proposed by Morris-Thorne \cite{Morris:1988cz}. It reads
\begin{equation}
ds^2=e^{2\Phi(r)}dt^2+\left(1-\frac{b(r)}{r}\right)^{-1}dr^2+r^2(d\theta^2+\sin^2\theta d\phi^2) 
\label{line-element}
\end{equation}
where $r,\,\theta$ and $\phi$ are the spherical coordinates, $\Phi(r)$ and $b(r)$ are two arbitrary functions of radius, $\Phi(r)$ is called the redshift function since it determines the gravitational redshift, and $b(r)$ determines the spatial shape of the wormhole we called the shape function. The traversable wormholes with three-form fields solutions for $\Phi(r)$ and $b(r)$ and its relevant quantities will be solved in the next section. Specifically, we will first use the wormhole line element to specify the ansatz of the three-form field at this moment.

According to Eqs.(\ref{3-form-dual}) and (\ref{3-form-dual-comp}), the nonzero components of three-form fields are given by
\begin{equation}
\begin{split}
A_{t\theta\phi}=A_{\theta\phi t}=A_{\phi t \theta}=-A_{t\phi\theta}=-A_{\theta t\phi}=A_{\theta t\phi}=-A_{\phi\theta t} =\sqrt{g\left(\frac{b(r)}{r}-1\right)}\zeta(r)
\end{split}    
\end{equation}
and $A^{2}$ is given by
\begin{equation}
A^{2}\equiv A^{\alpha\beta\gamma}A_{\alpha\beta\gamma}=-6\zeta(r)^{2}.
\label{A-squre}
\end{equation}
Having uses Eqs.(\ref{3-form-stength}) and (\ref{3-form-dual}), we can re-write the kinetic term of the three-form fields as
\begin{equation}
-\frac{1}{48}F^{2} = -\frac{1}{48}F^{\alpha\beta\gamma\delta}F_{\alpha\beta\gamma\delta} =\frac{1}{2}(\nabla_{\mu}B^{\mu})^2 =-6\Upsilon(r),      
\end{equation}
where the function $\Upsilon(r)$ with the traversable wormhole line element in Eq.(\ref{line-element}) reads
\begin{equation}
  \Upsilon=4\left( 1-\frac{b}{r}\right)\left[\zeta\left(\Phi^{\prime}+\frac{2}{r}+\zeta^{\prime}\right)\right]^{2}. 
  \label{upsilon} 
\end{equation}
We have completed a general form of the three-form field solutions in the spherical symmetric background. This will be used to construct the traversable wormholes in the next section.

\subsection{Traversable wormhole with three-form fields and and its energy conditions}
In the present section, we continue to derive the EFE with the three-form field in order to construct the traversable wormholes with the three-form field solution. More importantly, we also include an ordinary matter term, $\mathcal{L}_m$, into the gravitational action in Eq.(\ref{action}) and the ordinary matter term is considered as an anisotropic fluid in this work. Taking a standard variation of the action with respect to the metric tensor, $g_{\mu\nu}$ and setting $\kappa^{2}\equiv 8\pi G=1$, one gets
\begin{equation}
G_{\mu\nu}= {T}{^{\textrm{eff}}_{\mu\nu}},
\label{EFE-eff}
\end{equation}
where the effective energy-momentum tensor defined by
\begin{equation}
T^{\textrm{eff}}_{\mu\nu}=T^{(A)}_{\mu\nu}+T^{(m)}_{\mu\nu},
\label{EMT-eff} 
\end{equation}
where an $(A)$ superscript refers to the three-form $A_{\alpha\beta\gamma}$ and an $(m)$ denotes an ordinary matter of the energy-momentum tensor, respectively, while the energy-momentum tensor of the anisotropic fluid is given by $T_\nu^{(m)\,\mu} = \big(-\rho_m,\,-\tau_m,\,p_m,\,p_m \big)$.

Using the metric tensor in Eq.(\ref{line-element}), the energy-momentum tensor of the three-form, $T^{(A)}_{\mu\nu}$ in Eq.(\ref{EMT-3-form}) can be calculated for diagonal the components as
\begin{eqnarray}
{T}^{(A)}{^t_t}&=&-\rho_{A}=-\frac{1}{8}\Upsilon-V+\zeta \frac{\partial V}{\partial \zeta},
\label{rhoA}\\
{T}^{(A)}{^r_r}&=&-\tau_{A}=-\frac{1}{8}\Upsilon - V,
\label{tauA}\\
{T}^{(A)}{^\theta_\theta}&=&{T}^{(A)}{^\phi_\phi} =p_{A} =-\frac{1}{8}\Upsilon - V+\zeta \frac{\partial V}{\partial \zeta}. 
\label{pA}
\end{eqnarray}
From a definition of the effective energy-momentum tensor in Eq.(\ref{EMT-eff}), we can calculate the EFE in Eq.(\ref{EFE-eff}) for each diagonal component. One finds
\begin{eqnarray}
&&\frac{b^{\prime}}{r^2} = \rho_{\textrm{eff}} = \rho_{m}+\rho_{A},
\label{EFE-tt}\\
&& 2\left(1-\frac{b}{r}\right)\frac{\Phi^{\prime}}{r} -\frac{b}{r^3} = \tau_{\textrm{eff}} = \tau_{m}+\tau_{A},
\label{EFE-rr}
\\
&& \left(1-\frac{b}{r}\right)\left(\Phi^{\prime\prime}-\frac{b^{\prime}r-b}{2r(r-b)}\Phi^{\prime}+(\Phi^{\prime})^{2}+\frac{\Phi^{\prime}}{r}-\frac{b^{\prime}r-b}{2r^2(r-b)} \right) = p_{\textrm{eff}} =p_{m}+p_{A}. 
\label{EFE-phiphi}
\end{eqnarray}
According to the conservation of the effective energy-momentum tensor, $\nabla_{\mu}{T}^{(\rm eff)\,\mu}_\nu=0$, we obtain the equation of motion of the effective energy-momentum tensor as 
\begin{equation}
\tau^{\prime}_{\textrm{eff}}+\frac{2}{r}(\tau_{\textrm{eff}}+p_{\textrm{eff}})+\Phi^{\prime}(\tau_{\textrm{eff}}-\rho_{\textrm{eff}})=0 .
\label{conserv-EMT}
\end{equation}
Substituting all components of the effective energy-momentum tensor from Eqs.(\ref{EFE-tt}), (\ref{EFE-rr}) and (\ref{EFE-phiphi}) into EFE in Eq.(\ref{EFE-eff}), we find
\begin{eqnarray}
&&\zeta\left[r\Phi^{\prime}\left(b^{\prime}-\frac{b}{r}\right)+4+2b^{\prime}+2r^{2}\Phi^{\prime\prime}\left(\frac{b}{r}-1\right)-6\frac{b}{r}\right]
\nonumber\\
&&\qquad\, +\,2r^{2}\frac{\partial V}{\partial \zeta}+\zeta^{\prime}r\left[3\frac{b}{r}-4+b^{\prime}+2\Phi^{\prime}r\left(\frac{b}{r}-1\right) \right]
+2\zeta^{\prime\prime}r^{2}\left(\frac{b}{r}-1\right)=0 . 
\label{EOM}
\end{eqnarray}
In practice, it is very difficult to solve the exact solutions of the wormholes because of the unknown functions such as $\Phi$, $b$, $\rho_m$, $\tau_m$, $p_m$, $\zeta$ and $V$. Therefore, we need to assume a specific form of the redshift function, $\Phi(r)$ and shape function, $b(r)$ which are compatible with all requirements of the traversable wormhole solutions. More importantly, it has been demonstrated in the Ref.\cite{Barros:2018lca} that the following specific forms of the functions $\Phi(r)$, $b(r)$ and $\zeta(r)$ are consistent to the EFE in Eq.(\ref{EFE-eff}) with the line element in Eq.(\ref{line-element}) as well. The specific forms of the solutions in this work are taken from Refs.\cite{Lobo:2005us,Lobo:2008zu,Capozziello:2012hr}. The functions $\Phi(r)$ and $b(r)$ can be parameterized as
\begin{equation}
\Phi(r)=\Phi_0 \left( \frac{r_0}{r}\right)^{\alpha}  \quad \textrm{and} \quad b(r)=r_0\left(\frac{r_0}{r}\right)^{\beta}  
\label{ansazt-phi-b-z}    
\end{equation}
where $\alpha, \beta,$ and $\Phi_0$ are dimensionless and the ranges of the model parameters are assumed to be $\beta > -1, \alpha > 0$ and $\beta > 0$. $\Phi(r)$ and $\zeta(r)$ are also dimensionless function, whereas the $b(r)$ and $r_0$ are quantities with dimension of length.

In this work, we use a specific form of the three-form field potential as \cite{Barros:2018lca},
\begin{eqnarray}
V(r)&=& \zeta^2 + C\,. 
\label{V}
\end{eqnarray}
where $C$ in an integration constant. The author of Ref.\cite{Barros:2018lca} has shown that the relevant energy conditions for the three-form wormhole are satisfied by using the the form of $\Phi(r)$, $b(r)$ and $V(r)$ in Eqs.(\ref{ansazt-phi-b-z}) and (\ref{V}), respectively, with the following sets of the parameters: 
\\
$\bullet$ Case 1 
\begin{eqnarray}
\Phi_0=-1,\quad \alpha=1,\quad \beta=1, \quad C = -0.1\,.
\label{para1}
\end{eqnarray}
\\ 
$\bullet$ Case 2  
\begin{eqnarray}
\Phi_0=-2 ,\quad \alpha=1,\quad \beta=-1/2, \quad C = 0\,,
\label{para2}
\end{eqnarray}

The authors of Ref.\cite{Barros:2018lca} showed that all two cases are the best parameter sets that produce results compatible with the relevant energy conditions. Therefore, they are suitable to be used to construct wormhole solutions. Then, we employ above two cases of the parameters to study gravitational lensing of the three-form wormholes in the next section.

\section{Effective potential and Deflection angle of the three-form wormhole}\label{lens}
In order to study the trajectory of the photon in the three-form wormholes background, the effective potential from the geodesic equation plays the major role leading to analysis of the deflection angle of the photon trajectory. Detailed derivations of the relevant quantities for deflection angle are given in the standard framework and they can be found in general textbooks in GR \cite{Carroll:2004st,Hartle:2003yu}. Then we will not repeat them here. We aim to derive in this section the analytical form of the effective potential and deflection angle of light of wormholes with the three-from fields. We study the numerical calculations in the next chapter. 

In the previous section, we have reviewed how to construct wormholes solutions supported by three-form fields. In this section, we use the redshift function $\Phi (r)=\Phi_0(\frac{r_0}{r})^{\alpha}$ and the shape function $b(r)=r_0(\frac{r_0}{r})^{\beta}$ that can be used to construct wormhole solutions as mentioned previously. We divide the analysis of the effective potential from the geodesic equation and the deflect angle in this section into two cases with two sets of parameters in Eqs.(\ref{para1},\ref{para2}) as shown in the previous section. In addition, we consider different values of $\alpha$, $\beta$ and $\Phi_0$ relevant to the calculations to determine the effective potential as well as the deflection angle.

\subsection{Case I : $\alpha=1$, $\beta=1$ and $\Phi_0=-1$} 
We set $\Phi_0=-1$, $\alpha=1$ and $\beta=1$ leading to $\Phi(r)=-r_0/r$ and $b(r)=r_0^2/r$, respectively. From the standard approach in GR as done in Refs.\cite{Shaikh:2018oul,Shaikh:2019jfr,Godani:2021aub}, the effective potential, $V$, is defined by 
\begin{eqnarray}
V =\frac{e^{-\frac{2r_0}{r}}}{r^2}\,,\qquad \text{with}\quad r^2=l^2+r_0^2\,,
\end{eqnarray}
where $l$ is the proper radial coordinate of the travesable wormholes. With the parameters given in case 1, we find 
\begin{equation}
V=\frac{1}{l^2+r_0^2}\,e^{-\frac{2r_0^2}{\sqrt{l^2+r_0^2}}}. \label{V1}
\end{equation}
We will plot potential $V$ with respect to proper radial coordinate, $l$, in the section \ref{subsection:4.2.1}. The numerical results will be displayed in the next section.

According to the standard calculations (see Refs.\cite{Shaikh:2018oul,Shaikh:2019jfr,Godani:2021aub} for detail derivations), the deflection angle with the parameter set given in case 1 can be obtained as
\begin{equation}
\alpha(u)=2\int_{r_{\rm tp}}^{\infty}dr\,\frac{e^{-\frac{r_0}{r}}}{r^2\sqrt{\left(1-\frac{r_0^2}{r^2}\right)\left(\frac{1}{u^2}-\frac{e^-\frac{2r_0}{r}}{r^2}\right)}}-\pi. \label{de1}
\end{equation}
In addition, the impact parameter, $u$, of the case 1 is given by
\begin{equation}
u=r_{\rm tp}\,e^\frac{r_0}{r_{\rm tp}}\,.
\end{equation}
Moreover, the turning point of the effective potential, $r_{\rm tp}$ is defined by
\begin{equation}
r_{\rm tp}=-\frac{r_0}{W\big(\frac{r_0}{u}\big)}  
\end{equation}
where $W$ function is the Product-Log function. The numerical results of the deflection angle will be depicted and discussed in the next section.

\subsection{Case II : $\alpha=1$, $\beta=-\frac{1}{2}$ and $\Phi_0=-2$}
For the case 2, we fix $\Phi_0=-2$, $\alpha=1$ and $\beta=1$ to obtain $\Phi(r)=-2r_0/r$ and $b(r)=\sqrt{r_0 r}$, respectively. Having used the definition of the effective potential as shown previously, one can write the expression of the effective potential, $V$, as 
\begin{equation}
V=\frac{1}{l^2+r_{0}^2}e^{-\frac{4r_0}{\sqrt{l^2+r_0^2}}}. \label{V2}
\end{equation}
The numerical results of the potential $V$ with respect to proper radial coordinate $l$ will be plotted and the graph is depicted in Fig.(\ref{fig:V14all2}) in the next section. With the same procedure previously, the deflection angle of case 2 parameters is given by 
\begin{equation}
\alpha(u)=2\int_{r_{\rm tp}}^{\infty}dr\,\frac{e^{-\frac{2r_0}{r}}}{r^2\sqrt{\left(1-\frac{r_0}{r\sqrt{\frac{r_0}{r}}}\right)\left(\frac{1}{u^2}-\frac{e^-\frac{4r_0}{r}}{r^2}\right)}}-\pi .\label{de2}
\end{equation}
The impact parameter, $u$, of case 2 reads
\begin{equation}
u=r_{\rm tp}e^\frac{2r_0}{r_{\rm tp}}\,.
\end{equation}
In addition, the turning point, $r_{\rm tp}$ for case 2 reads 
\begin{equation}
r_{\rm tp}=-\frac{2r_0}{W\big(-\frac{2r_0}{u}\big)} ,    
\end{equation}
The numerical plot of the deflection angle with respect to impact parameter will be depicted in the next section. In this section, we have computed the effective potential, $V$, and the deflection angle, $\alpha$, for the the photon trajectories in the spacetime back ground of the three-form wormholes. The numerical results and some physical implication will be presented in the next section.

\section{Numerical results}\label{numer-results}
In this section, we will present the numerical results for the effective potential and the deflection angle by using the parameters given in case 1 and case 2 where they are satisfied the NEC and WEC for the three-form wormholes as demonstrated in Ref.\cite{Barros:2018lca}. 

\subsection{Effective potential}\label{subsection:4.2.1}
We start with a numerical study of the effective potential in Eqs.(\ref{V1}) and (\ref{V2}) for parameters of case 1 and case 2, respectively.  
\begin{figure}[h!]
    \centering
    \includegraphics[width=12cm]{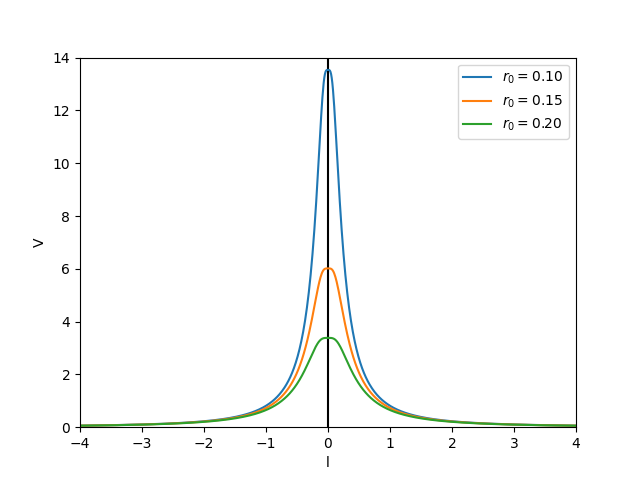}
    \caption{Plot for effective potential $V$ as a function of radial coordinate $l$ with different values of $r_0$. In this figure, we have used the parameters of case 1, $\Phi_0=-1,\alpha=1$ and $\beta=1$. with the ansatz of $\Phi(r)$ and $b(r)$ in Eq.(\ref{ansazt-phi-b-z}). }
    \label{fig:V14all1}
\end{figure}
\begin{figure}[h!]
\centering
\includegraphics[width=12cm]{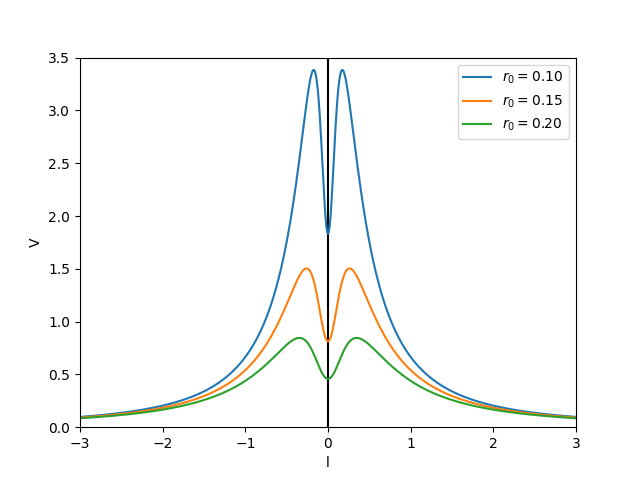}
\caption{Plot for effective potential $V$ as a function of radial coordinate $l$ with different value of $r_0$. In this case 2, we have considered $\Phi_0=-2,\alpha=1$ and $\beta=-1/2$ with the ansatz of $\Phi(r)$ and $b(r)$ in Eq.(\ref{ansazt-phi-b-z}).}
\label{fig:V14all2}
\end{figure}
Before discussing the numerical of the effective potential of the photon trajectory in the three-form wormholes background, it is worth considering the photon moving in the potential until reaching the “turning point" where $d\widetilde{V}/dr=0$. Here $\widetilde{V} \equiv L^2\,V$ with $L$ is the angular momentum of the photon circular motion. The constant radius $r_{\rm ph}$ is the constant radius of the photon motion in the circular orbit. The circular orbit of the photon occurs when
\begin{equation}
\frac{2e^{2\Phi(r_{\rm ph})}\Phi^\prime(r_{\rm ph})}{r_{\rm ph}^2}-\frac{2e^{2\Phi(r_{\rm ph})}}{r_{\rm ph}^3}=0.
\end{equation}
The radius of the photon circular orbits is given by
\begin{equation}
r_{\rm ph}=\frac{1}{\Phi^\prime(r_{\rm ph})}.    
\end{equation}
The radius $r_{\rm ph}$ is known as the photon sphere radius occurring at strong gravity region. In general, the conditions for examining the photon sphere radius are written as\\
\begin{equation}
\tilde{V}(r_{\rm ph})= E^2 \,, 
\quad \frac{d\tilde{V}}{dr}\bigg|_{r=r_{\rm ph}}=0 \,, 
\quad \frac{d^2\tilde{V}}{dr^2}\bigg|_{r=r_{\rm ph}}<0 \,,  
\label{Vcon}
\end{equation}
where $r_{\rm ph}$ and $E$ are the photon sphere radius corresponding to the maximum of the effective potential and the energy of the photon, respectively.

We have plotted effective potential for different values of $r_0$. As a result, we found that the numerical results in Fig.(\ref{fig:V14all1}) are compatible with the conditions (\ref{Vcon}) for all values of $r_0$ in the plots.

The light deflection occurs when the turning point of the photon approaches the throat $r_{\rm tp}=r_0$ of the wormholes. This implies the occurrence of the bending of light due to the photon sphere at the throat of wormholes \cite{Godani:2021aub}. In case 2, as a result, Fig.(\ref{fig:V14all2}) shows the maximum value of $\tilde{V}$ corresponding to the photon sphere radius, $r_{\rm ph}$. The light with turning point $r_{\rm tp} > r_{0}$ always take a turn outside the photon sphere. This implies that light diverges at the photon sphere outside the throat as the interpretation from Ref.\cite{Shaikh:2019jfr}.

\subsection{Deflection angle}\label{subsection:4.2.2}
\begin{figure}[h!]
\centering
\includegraphics[scale=0.5]{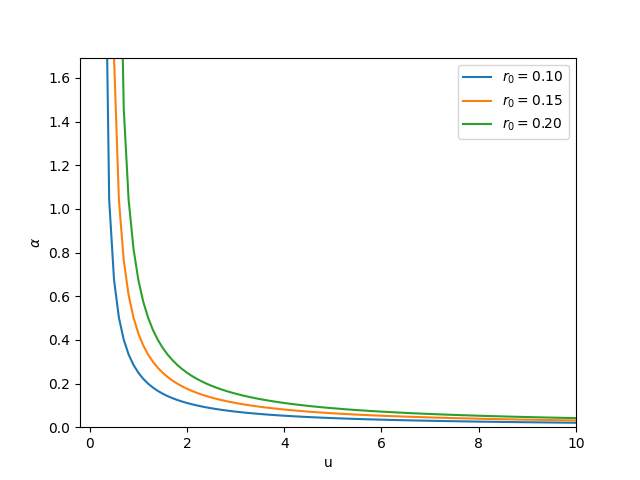}
\caption{Plot deflection angle $\alpha$ with respect to impact parameter $u$ for three different values of $r_0$ by using the ansatz of $\Phi(r)$ and $b(r)$  and the parameter set in case 1 in Eqs.(\ref{ansazt-phi-b-z}) and (\ref{para1}), respectively.}
\label{fig:deflec1}
\end{figure}
\begin{figure}[h!]
\centering
\includegraphics[scale=0.5]{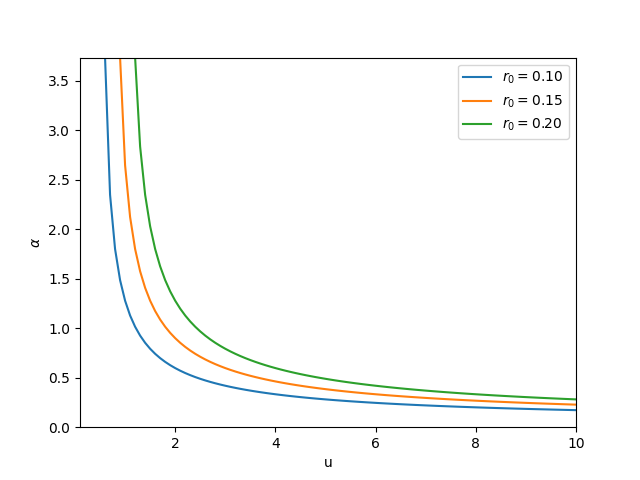}
\caption{Plot deflection angle $\alpha$ with respect to impact parameter $u$ for three different values of $r_0$ by using the ansatz of $\Phi(r)$ and $b(r)$ and the parameter set in case 2 in Eqs.(\ref{ansazt-phi-b-z}) and (\ref{para2}), respectively.}
\label{fig:deflec2}
\end{figure}
We have plotted numerical integration of the deflection angle from Eqs.(\ref{de1}) and (\ref{de2}) with respect to the impact parameter. Fig.(\ref{fig:deflec1}) and  Fig.(\ref{fig:deflec2}) show the divergence of $\alpha$ as the $u$ increase. the deflection of light occurs when the impact parameter approches the critical value $u\rightarrow u_{0}$. It is diverging to infinity corresponding to the throat radius $r_0$. When the throat of wormhole $r_0$ is large $\alpha$ is diverging slowly.    
We close this section by discussing the implications of the gravitational lensing created by the three-form wormholes. In case 1, on the one hand, the numerical results from Fig.(\ref{fig:V14all1}) and (\ref{fig:deflec1}) show that the potential, $tilde{V}$, has the maximum value at the throat and deflection angle of the photon getting diverge at the throat. This implies that the photon sphere can be used to figure out the throat of the three-form wormholes. On the other hand, the numerical results of case 2 shown in Fig.(\ref{fig:V14all2}) and (\ref{fig:deflec2}) show that the potential $\tilde{V}$ has the maximum value at the photon sphere radius. This means that the light bending occurs at the photon sphere outside the throat. In addition, the results of case 2 tell us that the observers will see two sets of Einstein's relativistic rings at and outside the throat. Comparing the numerical results of the deflection angle from two case studies, the solution from case 1 with $\Phi_0=-1$ has the deflection angle less than that of case 2 with $\Phi_0=-2$, meaning that the coefficient of the redshift function, $\Phi_0$, plays an important role in of the deflection angle analysis. Moreover, the deflection angles of both sets of the parameters from Fig.(\ref{fig:deflec1}) and (\ref{fig:deflec2}) imply that the larger wormholes produce the bigger deflection angles. 

\section{Conclusions}\label{conclusion}
In this work, we focus on the study of gravitational lensing from the traversable wormholes with the three-form fields. We used specific solutions to construct the wormhole solutions. We employ two sets of parameters and specific form of the three-form potential that are given by \cite{Barros:2018lca} which are most suitable to construct the energy densities and NEC profiles of the three-form fields and the matter source which does not violate the energy conditions. The null geodesics condition is used to find the geodesics of the wormhole metric to obtain effective potential. We have used the ansatz of the redshift and the shape functions in Eq.(\ref{ansazt-phi-b-z}) with the parameters of case 1 and case 2 to explain the existence of the photon sphere occurring in the wormhole background. As the results present in the present work, the geometrical shapes of the wormholes can be represented by using the observed photon sphere. Moreover, the Einstein's ring which is an useful physical observable in astronomy and cosmology can be used to probe the existence of the wormholes, see for example Refs.\cite{Tsukamoto:2012xs,Shaikh:2018oul}. In this work, we have provided an useful information that can be developed to predict the Einstein ring in the three-form wormholes. In addition, the gravitational signal of the wormholes has been drawn a lot of attention due to the active research fields in the gravitational wave observations. We will leave this interesting topic for the future work. 

\section*{Acknowledgement}
D. Samart is financially supported by the Mid-Career Research Grant 2021 from National Research Council of Thailand under a contract No. N41A640145. P. Channuie acknowledged the Mid-Career Research Grant 2020 from National Research Council of Thailand (NRCT5-RSA63019-03) and the National Science, Research and Innovation Fund (NSRF) with grant No.R2565B030.


\bibliography{ref.bib}


\end{document}